# SECURITY RISKS IN MECHANICAL ENGINEERING INDUSTRIES

Karen Benson[1] and Syed (Shawon) M. Rahman, Ph.D.[2]

[1]Information Assurance and Security, Capella University, Minneapolis, USA

KBenson1@capellauniversity.edu

[2]Assistant Professor, University of Hawaii-Hilo, Hilo, USA
and Adjunct Faculty, Capella University, Minneapolis, USA

SRahman@Hawaii.edu

## Abstract

*Inherent in any organization are security risks and barriers that must be understood, analyzed, and minimized in order to prepare for and perpetuate future growth and return on investment within the business. Likewise, company leaders must determine the security health of the organization and routinely review the potential threats that are ever changing in this new global economy. Once these risks are outlined, the cost and potential damage must be weighed before action is implemented. This paper will address the modern problems of securing information technology (IT) of a mechanical engineering enterprise, which can be applied to other modern industries.*

## 1. INTRODUCTION

Established in 1975, Bynsen and Coryn (B&C), Inc. is a mechanical engineering company which builds, installs, maintains, and designs large HVAC networks for hospitals, storage area networks, universities, and mega-structures throughout the world. B&C Inc. is an enterprise which has grown from 30 workstations in a peer-to-peer network to a multi-tiered Active Directory Domain with three virtual private network (VPN) tunnels and five remote sites. The company has transitioned from a fledgling direct subscriber line (DSL) network operated by an employee during afterhours to a full-throttle local area network (LAN) with a bundled 3-T1. Unfortunately, this rapid growth has left many security holes and threats to the enterprise to which many major stake holders have voiced concerns. This problem inhibits the company's goal of becoming a global partner in the mechanical engineering field. Using this particular case study, this paper will explore the methods necessary to secure the IT services of a mechanical engineering corporation, which are applicable to other areas of business and security.

## 2. CASE STUDY: B&C INC. MECHANICAL CONTRACTORS

Consider the following corporate site layout as portrayed in Figure 1. The main office, Alpha, houses the majority of the computer hardware for the company, including the Active Directory, FTP site, web site, and corporate databases. Consequently, this location is a prime target for a security breach. All sites are guarded with a SonicWall firewall, which also performs the tunneling for traffic between sites; however, the company limits the data going through the VPN tunnel because of the slow wide area network (WAN) link. Most data is moved via the FTP site.

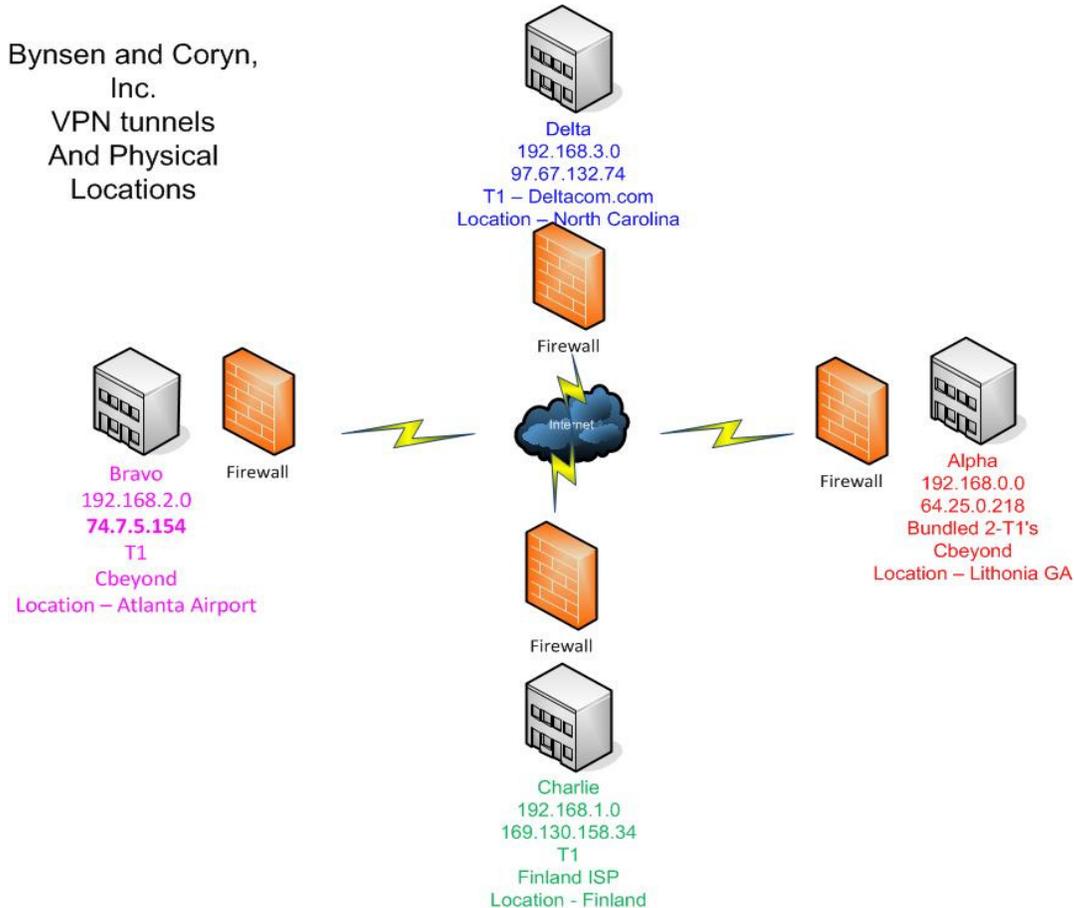

Figure 1. B&C, Inc. Sites and VPN structure

### 2.2 Functional Teams

This group is composed of the employees of B&C, Inc. who produce the goods and services to be sold to the public. These teams can be broken down into the following:

1. *Auto CAD* - These are the engineers and designers for the commercial HVAC networks.
2. *Workshop* – This group builds the million-dollar HVAC units designed by the engineers.
3. *Warehouse* – This group stores equipment necessary to build and maintain the HVAC units.

4. *Union Official* – This group that is a conduit between the management and the union employees who install and maintain all HVAC units for the customers.

5. *Project Management* – This group is composed of management personnel responsible for over-seeing the construction and final implementation of the HVAC units.

6. *Safety Inspector* – This safety group is responsible for training all employees in safe and secure practices while working on the job. They also inspect for possible OSHA infractions.

7. *Client Service and Maintenance* – This functional group is responsible for maintaining and servicing the HVAC units and components after the installation has occurred.

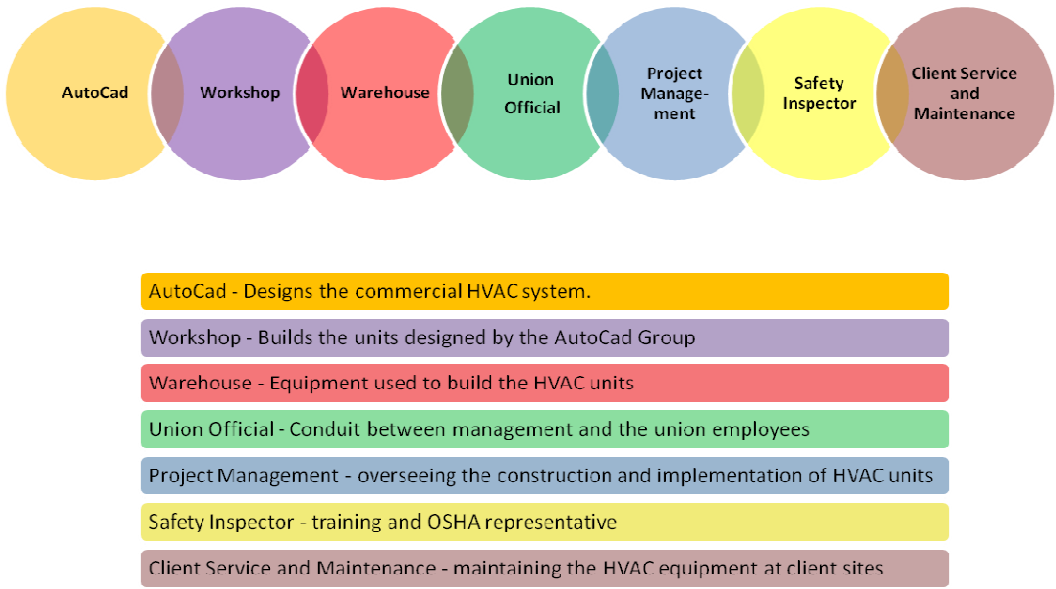

Figure 2. Functional Teams

## 2.3 Support Teams

These are the employees of B&C, Inc. who provide support to the functional teams, which allow for the execution of the day-to-day activities of the company. These teams can be broken down into the following:

1. *Accounting* – The accounting group provides all financial statements and interacts with an external CPA firm. In addition to the company's financial burden, the accounting team manages the pay of the company's employees, over 300 paychecks per week.

2. *Marketing* – The marketing group meets with advertising agencies and external engineering organizations to continually seek new building projects.

3. *Human Resources (HR)* – HR is responsible for all employee benefits to include hiring and firing of employees.

4. *Building Maintenance* – This group is responsible for the all functions of the campus buildings to include security, up-keep, cleaning, and build-outs.

5. *Information Technology (IT)* – The IT team is responsible for the design, installation, and maintenance of all electronics and the security of all data for the company.

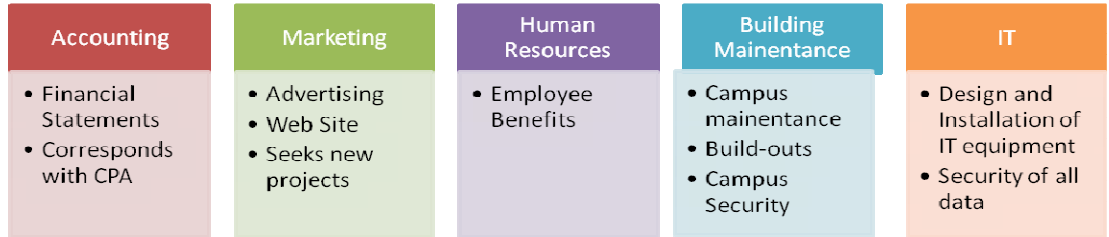

Figure 3. Support Teams

## 2.4 Corporate Structure

The Corporate hierarchy is structured as such:

1. President
2. Vice President
3. AVP of Operations – Over the Functional Teams
4. AVP of Network – Over the Support Teams
5. AVP of Service – Over all the field engineers who maintain the HVAC units

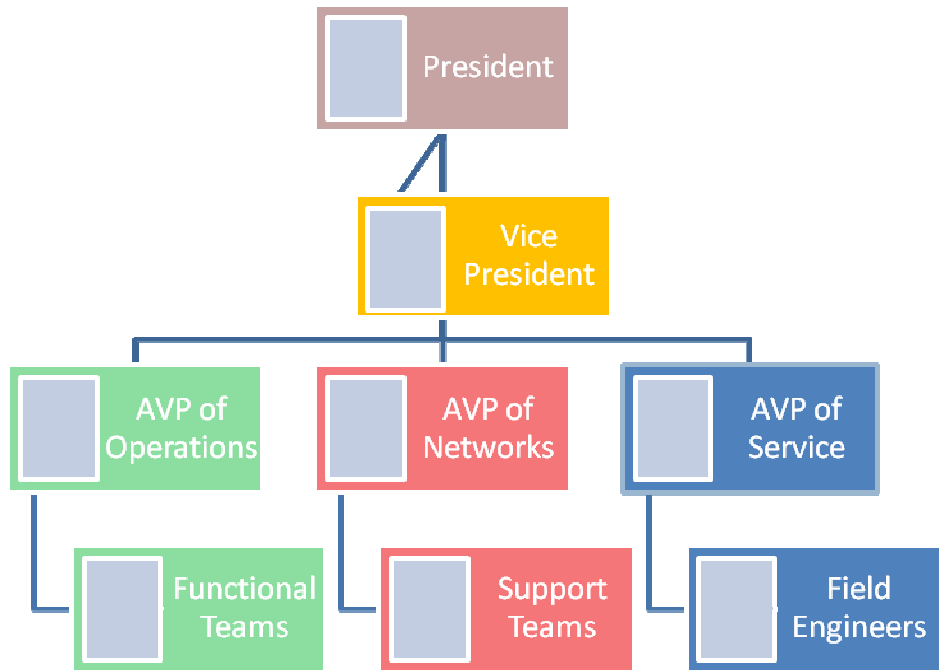

Figure 4. Corporate Structure

## 3.0 Security Posture

According to Lora Shinn, security posture is the overall security plan, which protects against internal and external threats to the enterprise network. Security posture is evident in the way the company deals with customer receipts, control of employee social security numbers, or how often the anti-virus software is updated. In other words, security posture is comprised of technical *and* non-technical policies, procedures, and controls (Shinn, 2010). Furthermore, Mike Murray suggests that there is a three-step approach to security posture assessment and resolution:

- Determine the data that competitors, thieves, and other malfeasants want to steal from your business or from partnering businesses. These could include credit card numbers, social security numbers, corporate assets, or even your business strategies for the next six months (Murray, 2011).

- Determine how thieves might acquire the data. This step may require a consultant or an in-house expert in risk management. A high-quality inspection will assess the shortfalls within a particular data set, whether in the IT or physical world. "We aptly call it information security, not just technology security." (Murray, 2011).

- Install controls to prevent theft, at a "palatable" cost. Your response may depend upon variables such as your business's financial situation and the actual likelihood of compromised data (Murray, 2011).

Because B&C, Inc. designs, builds, and maintains the HVAC units for top security clients, IT security is a top priority of the enterprise. The IT infrastructure must always be cognizant of inerrant security threats from inside and outside the organization. Mitigation of possible threats consists of an incidence response plan, a disaster recovery plan, and a continuance plan which documents and enforces policies among all personnel both in the offices and the field. The IT staff must be constantly aware of new risk potentials and how to effectively reduce and eliminate their danger to the organization. Balancing risk and cost is part of the management process of the IT personnel.

## 4.0 ANALYSIS OF ENTERPRISE SYSTEM WEAKNESS

It is impossible to analyze the IT security of an enterprise by looking solely at the computers and network. It must be examined in its entirety with regards to a global view of company personnel policies, building infrastructure, computer software, and computer hardware. A security analysis must first start with a comprehensive study of the enterprise, its business function, purpose, and financial data and then must shift focus to a micro analysis of the symbiotic parts (Whitman, M. and Mattord, H., 2010). Since one of the installation clients of B&C, Inc. is a highly secure data storage area network, the enterprise security at B&C, Inc. must be enforced. The macro view of the enterprise system security can be compartmentalized into four distinct parts: physical structure, computer software, IT network, and company personnel. Similarly, each of these compartments must be analyzed as to the possible weaknesses and intrusions that can occur, which would compromise the enterprise security.

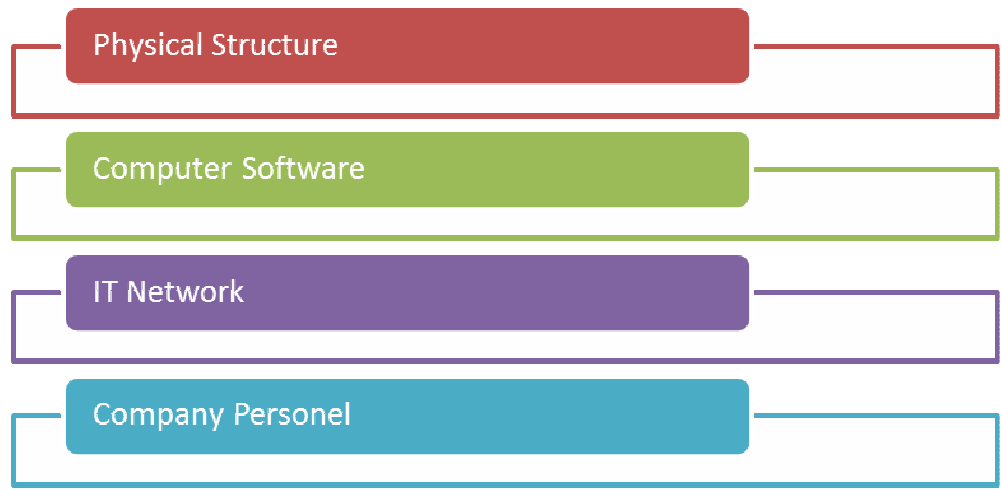

Figure 5. Areas of Vulnerability

## 4.1 Physical Structure

Each security analysis must begin with the strength and weaknesses of the physical structure, without which there is no data integrity to ensure. Being a Butler-Building, which was constructed in the 1970's, it is very prone to tornado and high wind threats. Until recently, a climate shelter for employees and data did not exist. Also, windows at ground level can be easily vandalized. Even with warnings from the IT team, the management continues to locate the accounting server by the window most vulnerable to a break-in. Fortunately, the building has a tight security fence and card gate. Security cameras are located on the inside and outside and have a monitoring system present at all times. Accordingly, there is a security alarm which utilizes a system whereby each employee has a unique entry code. In this manner, it can record the times of entry and exit of each employee; however, on occasion, someone's key is compromised, which requires the reset of the entry code. All wiring to include the voice and data are vulnerable to exposure due to open cables entering the building. The DMark voice and data router are located on an unprotected outside wall, which easily could be attacked or disconnected. Likewise, the building is not protected by vehicle barriers and arresters meaning that at any time an intruder could drive through the front door and steal computers, equipment, and data. Also, the physical building does not have a sprinkler system, which would be effective in saving lives and hardware in the event of a fire. Finally, the server room is not locked and is entirely vulnerable to inside attacks and theft.

## 4.2 IT Network

The IT Network is also an integral part of the macro view of B&C, Inc. security review. Consisting of a main office and 3 satellite offices, the network is a magnet for hackers. The remote offices are connected via an unsecured VPN tunnel by DSL connectivity. Similarly, the server closets in the remote offices are not locked and equipment could easily be hacked and stolen. Connecting all the sites are firewalls which use VPN tunnels, and each firewall has a secure password; however, the username on the firewall is 'admin' which is easily attacked. The firewall has Comprehensive Gateway Security; however, it has been disabled because the packet analysis causes the Internet to be slow. Also on the firewall is the wireless system which, due to the president's request, is bridging the Wireless LAN and the internal LAN. Anyone who knows the wireless password, can jump onto the LAN and view user's data. The FTP port 21 is also opened on the firewall. To date, the company does not have any FTP software which would eliminate the 'hammering' of the firewall by dictionary attacks. This means that every second, an intruder is using a dictionary to attack the FTP open port. Because the policy of management is 'ease of use', the IT network is wide open to attack by both theft and hackers.

## 4.3 Computer Software

Security of the IT software at B&C, Inc. is comparatively lacking due to the ease-of-use company policy. The passwords to all workstations are identical which creates a problem when an employee is terminated. Also at risk is the AutoCAD software which is not locked and is vulnerable to theft by employees. Microsoft software is automatically updated each night; however, if the employee turns off his computer, the workstation will not be updated. This is crucial since the Internet Explorer that is used, if not updated, can be a potential threat for viruses. In order to secure the accounting data, the Accounting Server is on its own local area network which does not go to the Internet or touch the company local area network. In theory, this should keep the software and data safe; however, since the server does not reach the network, it does not receive Microsoft updates or antivirus updates. At any time, an employee can put in a flash drive or CD which is virus infected which would bring down the accounting department. Similarly, each workstation is not locked down meaning that any user can download software or insert and install software from a CD or flash drive. This could comprise the network at B&C, Inc. within seconds leading to a crucial work shut-down with 250 union paychecks at stake.

## 4.4 Company Personnel

Most important, the leaders of knowledge organizations fully realize that their most important assets walk out the door every night (Huseman & Goodman, n.d.). The final and most important line of defense are the people in the organization. Granger (2001) describes social engineering as a hacker's clever manipulation of the natural human tendency to trust. Generally agreed upon as the weakest link in the security chain, the natural human willingness to accept someone at his or her word leaves many of us vulnerable to attack. Many experienced security experts emphasize this fact. No matter how many articles are published about network holes, patches, and firewalls, we can only reduce the threat so much, and then it is up to Maggie in accounting or her friend, Will, dialing in from a remote site, to keep the corporate network secured (Granger, 2001). At Bynsen and Coryn, Inc. there is a great vulnerability in the personnel aspect as there is no education as to the importance of IT security. An example occurred recently when a vendor asked an employee for the wireless password upon which the employee wrote it down on paper for the vendor and everyone at his company. There are no policies in place as to what documents need to be shredded, for instance, company hardware documentation or personnel manuals. Another example of a lack in policy is the use of laptops on vacation. Not aware of the security dangers, several of these pieces of equipment have been stolen because of employee negligence. This caused a panic and many man-hours of changing passwords and monitoring accounts for intrusion due to the theft. Finally, again due to lack of education, many employees do not see the inherent danger of letting family members utilize the work laptop for recreational purposes. Would you let your teen-age son learn to drive on the company car?  Probably not, and this is the way that it should be when company property is in the employee's care.

In any analysis of IT defense, the relationship of cost vs. security must be discussed and determined then policies must be written and enforced.  Personnel must be educated as to the policy and then sign-on to the security knowing that if the company fails because of a security breach, the employee will ultimately loose his/her job. By demonstrating leadership skills, the IT director/contractor can guide the security of a business entity into a successful going concern.

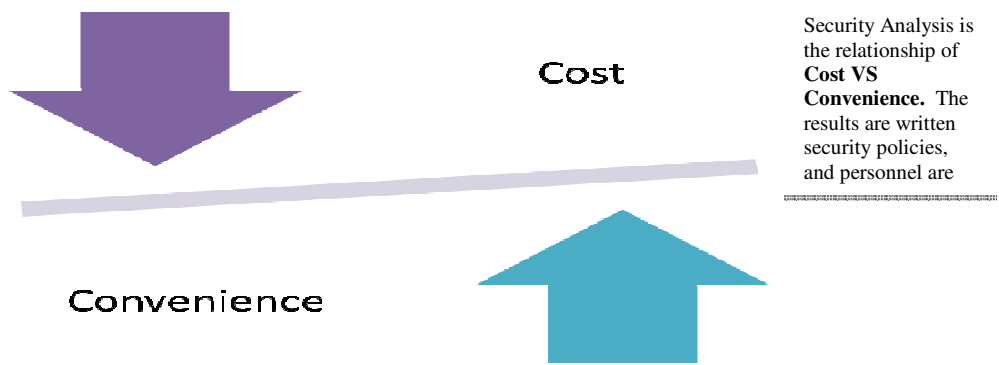

Figure 6. Cost VS. Convenience

## 5.0 ANALYSIS AND DESCRIPTION OF SECURITY REQUIREMENTS

A security analysis must first start with a comprehensive study of the enterprise, its business function, purpose, and financial data, then a more micro analysis of the symbiotic parts should be considered (Whitman, M. and Mattord, H., 2010). Security of the IT software at B&C, Inc. is comparatively lacking due to the ease-of-use company policy.  This is exemplified by the following observations from a detailed, on-site security analysis:

1. Password policy is non-existent

2. Computer software is vulnerable to theft both physical and license key

3. Computers are not automatically updated with Windows Software

4. Accounting server is by a window and susceptible to theft

5. Accounting server is not backed up

6. No policy on employees using flash drives and writeable CD's

7. No policy on employee downloading software to workstations

8. No policies in place for shredding of documents

9. Employees are free to use their business laptops for personal use

10. Critical data is not backed up offsite

11. FTP sites are vulnerable to 'hammering'

12. The Content Gateway Security System is turned off to allow for faster Internet access

13. Accounting offices are not locked

14. Remote offices are not secured

15. There is no inventory system for computers and company assets

16. There never has been a security audit of the company

17. There is no policy on employee termination and subsequent account login

## 5.1 Most Basic Security Threats - NISTIR

According to National Institute of Standards and Technology (NISTIR, 1993), the following statements are the ***most basic security threat***s that must be addressed in an organization:

1. An unauthorized user may attempt to gain access to the system.

2. An authorized user may attempt to gain access to resources for which he or she is not allowed access.

3. Security relevant actions may not be traceable to the individual associated with the event.

4. The system may be delivered, installed, or used in an unsecured manner.

5. Data transmitted over a public or shared data network may be modified either by an unauthorized user or because of a transmission error or other communication-related error.

6. Security breaches may occur because available security features are not used or are used improperly.

7. Users may be able to bypass the security features of the system.

8. Users may be denied continued accessibility to the resources of the system (i.e., denial of service).

## 5.2 Security Features and Assurances

The following are *security features and assurances* that are required to counter the above basic threats (NISTIR, 1993):

**Identification and Authentication**- Property access, as well as, computer access must require employee/user identification using user-id's that are adequately protected against fraud. For employees that have been terminated, the organization must have a secure method of ensuring that the terminated employee can not retrieve or alter information. Accordingly, strong password policies must be enforced (NISTIR, 1993).

**Access Control** – Access control determines what an authenticated user can do to a system. Two types of access control are considered here: system access and resource access (NISTIR, 1993). In order to achieve this rational, the identity of all users shall be authenticated before access is granted to any resources or system information. Similarly, all remote users must have a secure and strong password to access the main system from remote offices. If the user is denied access, the system needs to log this invalid entry for the system administrator's review.

**System Integrity** - Users expect to share computer resources without interference or damage from other users. This includes data, protection of software, firmware, and hardware from unauthorized modifications (whether deliberate or accidental), and control of operator and maintenance personnel actions. Used in accomplishing these tasks are job logs, audit functions, software updates, machine maintenance, and detection of communication errors (NISTIR, 1993).

**Data Integrity** - Users expect data to be entered and maintained in a correct, consistent state. This expectation applies to both user data and system data. This requires mechanisms that promote tracking of changes to resources, the protection of data against exposure, and unauthorized modification or deletion as it is transmitted and while it is stored. To ensure data integrity, the system administrator must enforce data encryption when communicating with remote sites, and read logs which indicate that a security breach has occurred (NISTIR, 1993).

**Reliability of Service** - Users expect a quantifiable and reliable level of service from a system. These mechanisms also allow prevention or limitation of interference with time-critical operations, and allow the system to maintain its expected level of service in the face of any user action threatening this level, whether the action is deliberate or accidental. This involves giving users disk quotas for data storage, uninterrupted power supplies for all hardware, and software backup systems (NISTIR, 1993).

**Product Documentation Assurance** –Documentation for users, administrators, and operators must be present to support the secure installation, operation, administration, and use of the product. The requirements for product documentation assurances are intended to ensure that security breaches do not occur because available security features are not used or are used improperly. This documentation will contain instructions on actions to be carried out if security is breached within the organization.

Also, contained in this documentation will be backup policies and procedures along with timetables to read and examine audit logs (NISTIR, 1993).

## 5.3 Expanding Zones of Vulnerability

According to Mar (2010), it is important as an IT director to be familiar with the zones of vulnerability that can attack an organization.

**Corporate LAN** – The emphasis here is on the centralized resources found in the Local Area Network. Included are DOS (denial of Service), viruses, worms and Trojan Horses (Mar, 2010).

**BackBone LAN** – The focus here is the distributed resources which occur on the Backbone LAN. These included unauthorized access, Connection Hijacking, and DoS (Mar, 2010).

**Branch Network** – The outer offices in an organization are part of the ever increasing vulnerability. In focus are electronic theft mechanisms, masquerading, and eavesdropping (Mar, 2010).

**Access Network** – Employees, Customers, and Partners are the outer edge of the Zone of Vulnerability. This includes PC Hijacking, NSP Congestion Attacks, and Virus Attacks (Mar, 2010).

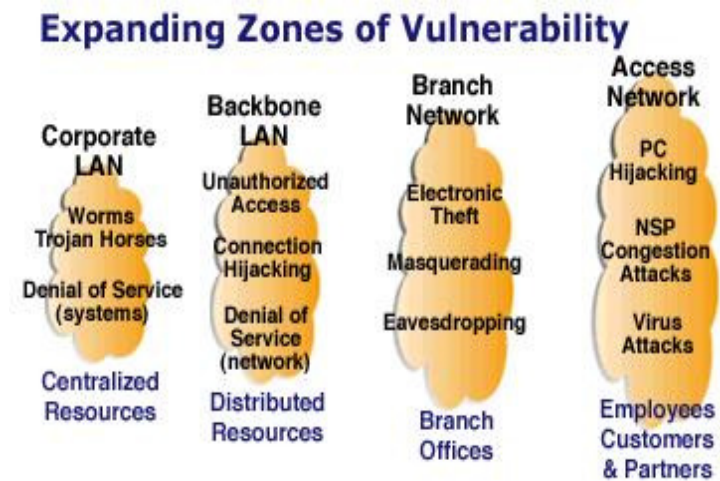

Figure 7. Expanding Zones of Vulnerability (Mar, 2010)

## 5.4 Security Policy

In order for an organization to coordinate and educate concerning the security of an organization, a *security policy* must be developed, analyzed, tested and taught to every employee within the concern. According to Whitman and Mattord (Whitman and Mattord, 2010), a quality information security program begins and ends with a policy. This policy explains the will of the organizations management in controlling the behavior of the employees.

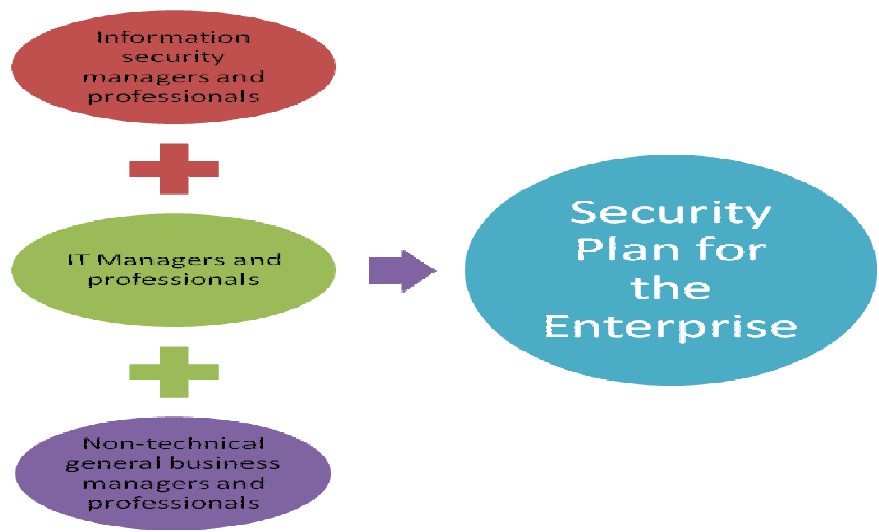

Figure 8. Security Plan Equation (Whitman and Mattord, 2010)

Security policies abound both on the Internet and in book form; however, the most basic policy statements should never conflict with the law, must be able to stand up in court if challenged, and must be supported and carried out by administration (Whitman and Mattord, 2010):. While a security policy must be tailored to the specific needs of an organization, the basic theme of the policy statement must include these principles (Whitman and Mattord, 2010):

1. All policies must contribute to the success of the organization
2. Management must ensure the adequate sharing of responsibility for proper use of information systems
3. End users of information should be involved in the steps of policy formulation

Also included in a security policy are penalties for unacceptable behavior and a definition of the appeals process. Included in this definition are visits to inappropriate web sites, compromising of company data, and breaching security within the organization whether accidental or intentional. This security policy must also delegate the responsibilities of auditing and ensuring that there is a 'watchdog', otherwise, the security policy is rendered useless. Most importantly, however, a security policy should indemnify the organization against liability for an employee's inappropriate behavior or illegal use of the system. Frequent updates and revisions must be made as the organization grows and changes with the technological times and economy. An outline for a security policy would be as follows (Whitman and Mattord, 2010):

1. Statement of purpose

2. Authorized uses
3. Prohibited uses
4. Systems management
5. Violations of policy
6. Policy review and modification
7. Limitations of liability

# 6.0 Security Plan for the Enterprise

The security of an organization must start with a basic security plan in which the ultimate defense goals and objectives are delineated. The process should involve three distinct groups of decision makers, or communities of interest (Whitman and Mattord, 2010):

1. Information security managers and professionals
2. IT managers and professionals
3. Non-technical general business managers and professionals

Working together, these communities of interest make collective decisions about how to secure the organization's information assets most effectively. Likewise, the importance of security can be visualized by understanding where the potential risks may occur in an entity. Whitman and Mattord (2010) describe these areas as:

1. **Physical security** -encompasses strategies to protect people, physical assets, and the workplace from various threats, including fire, unauthorized access, and natural disasters
2. **Operations security**- focuses on securing the organization's ability to carry out the operational activities without interruption or compromise
3. **Communications security**- encompasses the protection of an organization's communications media, technology, and content, and its ability to use these tools to achieve the organization's objectives
4. **Network security**- addresses the protection of an organization's data networking devices, connections, and contents, and the ability to use that network to accomplish the organization's data communication functions
5. **Confidentiality** - only those with sufficient privileges and a demonstrated need may access certain information.
6. **Integrity** - The integrity of information is threatened when it is exposed to corruption, damage, destruction, or other disruption of its authentic state.
7. **Availability** - Information that enables user access to information in a usable format without interference or obstruction.
8. **Privacy** - Information that is collected, used, and stored by an organization is intended only for the purposes stated by the data owner at the time it was collected.
9. **Identification** - An information system possesses the characteristic of identification when it is able to recognize individual users.
10. **Authentication** - Authentication occurs when a control proves that a user possesses the identity that he or she claims.
11. **Authorization** - After the identity of a user is authenticated, a process called authorization assures that the user (whether a person or a computer) has been specifically and explicitly authorized by the proper authority to access, update, or delete the contents of an information asset.

12. **Accountability**- Accountability of information exists when a control provides assurance that every activity undertaken can be attributed to a named person or automated process.

# 7.0 Current Risk Assessment of B&C, Inc. and Integration of the Security Plan

Each security analysis must begin with the strength and weaknesses of the physical structure, without which there is no data integrity to ensure. The current risk assessment can be prioritized into the following risks:

1. Butler Building that has no external security from 'smash and grab' theft occurrences. This could be remedied by an external tornado fence surrounding the building. Vehicle arresters should be place around the building, as well.  Also under consideration is to move the accounting server away from the floor length windows.

2. All wiring, including the voice and data are via exposed cables coming into the building.  The ISP voice and data router are located on an unprotected outside wall which could be easily attacked or disconnected.  This could be corrected by moving all the cabling back inside the interior server room. A lock and sign in sheet must be place on the server room door.

3. Remote offices using a VPN tunnel are not using SSL connections.  Also, the wireless security at the remote offices is not password protected. The security plan must have in place SSL VPN connections and all connections must have WEP wireless security with MAC address protection.

4. The firewall has Comprehensive Gateway Security; however, it has been disabled because the packet analysis causes the Internet to be slow. Also on the firewall is the wireless system which, due to the president's request, is bridging the Wireless LAN and the Internal LAN. Anyone who knows the wireless password, can jump onto the LAN and look at users' data. This would be disabled under a proper security plan.

5. The FTP port 21 is also opened on the firewall. To date, the company does not have any FTP software which would eliminate the 'hammering' of the firewall by dictionary attacks.  This means that every second, an intruder is using a dictionary to attack the FTP open port. The software WS_FTP would be implemented for proper protection.

6. The passwords to all workstations are identical which creates a problem when an employee is fired. A proper password protection policy must be implemented.

7. Microsoft software is automatically updated each night; however, if the employee turns off his computer, it will not be updated.  This is crucial since the Internet Explorer that is used, if not updated, can be a potential threat for viruses.  This would be corrected with a policy which states that all computers must be left on at night, but logged out.

8. Employees are using the 'auto-remember passwords' on their laptops.  If a laptop is stolen, all usernames and password are remembered which would enable a thief to log into corporate software. This problem could be remedied by a policy which states that no passwords can be store on a computer.

9. At any time, an employee can put in a flash drive or CD which is virus infected which would bring down the accounting department or steal data from the corporation.  This risk could be mitigated by not allowing any CD burning or flash drives in the building.

| Security Plan for the Enterprise | Security Weakness in B&C, Inc. | Security Correction Proposed for B&C, Inc. |
|---|---|---|
| Building Physical Assets | No External security<br>Exposed cabling system<br>Server room exposed | Install building arresters<br>Move DMark to server room<br>Install lock on server room door. |
| Operations Security | No Sprinkler system<br>No Uninterrupted Power Supply<br>No offsite backup<br>Accounting Server vulnerable | Install building-wide sprinkler<br>Each server will have a UPS<br>All backups will go to the 'cloud'<br>Move the Accounting server to a more secure location |
| Communications Security | VPN tunnels to outer offices not secure<br>FTP site unsecure<br>RDP protocol used to remote | Citrix will be used in the VPN tunnels<br>SFTP will be utilized on the FTP server<br>Citrix will be used instead of RDP protocol |
| Network Security | Wireless is not secured<br>Laptops are not secured<br>Critical workstations are not secured | Setup wireless WAP security with MAC filtering<br>Encrypt all data on the laptops and backup using the 'cloud'<br>All workstations that perform banking functions are not allowed to browse the Internet |
| Confidentiality | All passwords are known to all employees<br>IT is not informed of employee termination<br>All folders give users Full Control permissions | Implement a password policy<br>Implement an employee termination IT informed policy<br>Secure folders by determining what users need to access that folder |

| Security Plan for the Enterprise | Security Weakness in B&C, Inc. | Security Correction Proposed for B&C, Inc. |
|---|---|---|
| Integrity | All data is unencrypted | Encrypt data that is stored on the servers<br>Encrypt data that is stored on laptops |
| Availability | No backups are performed | Backup all servers to the 'cloud'<br>Backup all laptops to the 'cloud'<br>Perform test restores to ensure backups availability |
| Privacy | mployees use 'auto-remember' for passwords | Disallow the computer remembering passwords, especially to bank sites |
| Identification | Any employee can use the wireless<br>Any employee can enter the server room | Allow only authorized users to use the wireless<br>Disable the broadcast of the wireless<br>Lock the server room |
| Authentication | Emails have no authentication | Use asymmetric keys to encrypt emails to ensure that the user actually sent the email and not a hacker |
| Authorization | Web content filtering has been disabled to increase speed to the Internet | Enforce Comprehensive Gateway Security on the Firewall |
| Accountability | No training of personnel on security issues<br>No record keeping on security issues | Personnel will be trained routinely as to security issues<br>Security breaches will be recorded and used for training purposes |

Table 1. Table Showing Security Plan and Proposed Changes to Enhance B&C, Inc. Security

# 8.0 CONCLUSION

Securing an organization begins with knowledge of the specific industry, knowledge of the technology used in the organization, and knowledge of security practices. Careful examination of the organization sheds light on security weaknesses and red-flag precautions. From this information can be drawn a security policy that aids the employees to be cognizant of the importance of security and helps management build a defensive security position. Accordingly, security of an organization can not be mitigated or underestimated or the consequences to the company will be detrimental in both the short and long term. Also, it is necessary to show stock-holders and other financially interested parties that protection of the business and customer data is of utmost concern by the company leadership. With this security policy in focus at all times, organizations will have a greater chance of success in our current global economy. In observing the case study of B&C, Inc., one can see that it is critical to diagram the organization into Functional teams, Support teams, and Corporate Structure in order to fully understand and analyze the security vulnerabilities. Finding a solution to the security holes involves understanding the Security Posture and analyzing the Enterprise security weaknesses which include the physical structure, the IT network, computer software, and company personnel. Likewise, Security Features and Assurances are required to counter basic threats to the IT network. Finally, a security policy must be enforced and used for personnel training purposes to complete the protection circle.

**AUTHORS**

**Karen C. Benson** is a student at Capella University obtaining her Masters of Information Degree in information security. Karen has achieved the Microsoft Certified Systems Engineering certificate, as well as, the Microsoft Small Business Specialist certificate. Working as a network design engineer for Star Networks for 7 years, her job functions include: Voice over IP, Firewall, and Network Consultations, as well as, business systems and IT security consulting for small to medium sized enterprises. Karen is also the Cloud Computing expert for her company, Star Networks.

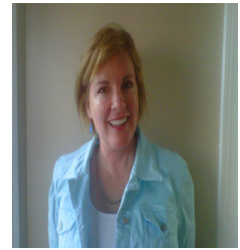

**Syed (Shawon) M. Rahman** is an assistant professor in the Department of Computer Science and Engineering at the University of Hawaii-Hilo and an adjunct faculty of information Technology, information assurance and security at the Capella University. Dr. Rahman's research interests include software engineering education, data visualization, information assurance and security, web accessibility, and software testing and quality assurance. He has published more than 50 peer-reviewed papers. He is a member of many professional organizations including ACM, ASEE, ASQ, IEEE, and UPE.

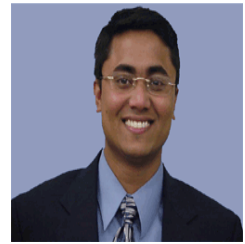